\documentclass[preprint,prl, superscriptaddress]{revtex4-2}
\usepackage{titlesec}%
\usepackage{subfigure}%
\usepackage{amsmath}%
\usepackage{amsfonts}%
\usepackage{amssymb}%
\usepackage{feynmp}%
\usepackage{graphicx}%
\usepackage{setspace}%
\usepackage{comment}%
\usepackage{dsfont}%
\usepackage{color}%
\usepackage{bbold}
\usepackage{tikz}
\usetikzlibrary{arrows,calc}
\usepackage[pdftex,colorlinks=true,linkcolor=blue,citecolor=blue]{hyperref}

\newcommand{\smode}{$s$ orbital\xspace}    
\newcommand{\pmode}{$p$ orbital\xspace}
\newcommand{\gbl}{gamma-butyrolactone\xspace}

\usepackage{xspace}
\usepackage{siunitx}

\titleformat{\paragraph}
{\normalfont\normalsize\bfseries}{\theparagraph}{1em}{}
\titlespacing*{\paragraph}
{0pt}{3.25ex plus 1ex minus .2ex}{1.5ex plus .2ex}

\begin{document}

\title{Photonic quadrupole topological insulator using orbital-induced synthetic flux}

\author{Julian Schulz\footnotetext[1]{These authors contributed equally}\footnotemark[1]}
\affiliation{Physics Department and Research Center OPTIMAS, TU Kaiserslautern, 67663 Kaiserslautern, Germany}
\author{Jiho Noh\footnotemark[1]}
\affiliation{Department of Mechanical Science and Engineering, University of Illinois at Urbana–Champaign, Urbana, IL 61801 USA}
\author{Wladimir A. Benalcazar}
\affiliation{Department of Physics, Princeton University, Princeton, New Jersey 08542, USA}
\author{Gaurav Bahl}
\affiliation{Department of Mechanical Science and Engineering, University of Illinois at Urbana–Champaign, Urbana, IL 61801 USA}
\author{Georg von Freymann}
\affiliation{Physics Department and Research Center OPTIMAS, TU Kaiserslautern, 67663 Kaiserslautern, Germany}
\affiliation{Fraunhofer Institute for Industrial Mathematics ITWM, 67663 Kaiserslautern, Germany}

\date{\today}

\maketitle

\textbf{Abstract}

The rich physical properties of multiatomic molecules and crystalline structures are determined, to a significant extent, by the underlying geometry and connectivity of atomic orbitals.
This orbital degree of freedom has also been used effectively to introduce structural diversity in a few synthetic materials including polariton lattices nonlinear photonic lattices and ultracold atoms in optical lattices.
In particular, the mixing of orbitals with distinct parity representations, such as $s$ and $p$ orbitals, has been shown to be especially useful for generating systems that require alternating phase patterns, as with the sign of couplings within a lattice.
Here we show that by further breaking the symmetries of such mixed-orbital lattices, it is possible to generate synthetic magnetic flux threading the lattice.
This capability allows the generation of multipole higher-order topological phases in synthetic bosonic platforms, in which $\pi$ flux threading each plaquette of the lattice is required, and which to date have only been implemented using tailored connectivity patterns.
We use this insight to experimentally demonstrate a quadrupole photonic topological insulator in a two-dimensional lattice of waveguides that leverage modes with both $s$ and $p$ orbital-type representations.
We confirm the nontrivial quadrupole topology of the system by observing the presence of protected zero-dimensional states, which are spatially confined to the corners, and by confirming that these states sit at the band gap.
Our approach is also applicable to a broader range of time-reversal-invariant synthetic materials that do not allow for tailored connectivity, e.g. with nanoscale geometries, and in which synthetic fluxes are essential.

\newpage
When studying materials and their physical properties, much emphasis is put on how atoms are combined to form molecules and crystalline structures through different orbital connections.
In particular, the richness of macroscopic properties in multiatomic molecules and crystalline structures is closely related to the way in which orbitals connect.
For example, a water molecule has an angled shape due to the hybridization of $s$ and $p$ orbitals, which in turn, causes the six-fold rotational symmetry of snowflakes. 
Similarly, the electronic properties of monolayer 2D transition metal dichalcogenides can be deliberately tuned over a wide range, in part, due to the interplay between $d$ orbitals on metal atoms and $p_{z}$ orbitals on chalcogen atoms \cite{Nanophotonics9Jing, NanoLetters10Splendiani, PRB83Kuc}.
In this context, the theory of ``topological quantum chemistry'' \cite{Nature547Cano} has greatly advanced the understanding of the intimate relationship between electronic orbitals and topological phases in crystalline structures and has led to the realization that topologically nontrivial materials are much more common than previously thought.

The ability of synthetic systems to replicate, and in many cases extend, the properties of chemical compounds and crystalline structures has recently become of great interest.
As in real materials, the orbital degree of freedom can also be incorporated into synthetic systems by using analogous wavefunctions with distinct nodal structures. 
Such synthetic multi-orbital systems have been demonstrated in polariton lattices \cite{PhysRevLett118Milicevic, PhysRevX9Milicevic}, photonic lattices \cite{PhysRevA100CaceresAravena, PhysRevA102CaceresAravena, PhysRevLett127GuzmanSilva, jorg_artificial_2020}, and ultracold atoms in optical lattices \cite{NatPhys7Wirth, NatPhys8SoltanPanahi, PhysRevA92Yin, PhysRevLett106Sun, NatPhys8Sun, NatCommun4Li}, with the possibility of negative \cite{NatCommun4Li, PhysRevA102CaceresAravena} and even complex-valued \cite{PhysRevA.93.033613,jorg_artificial_2020} coupling coefficients.
These features have made synthetic platforms well-suited to explore novel physics that typically are difficult to study in solid-state systems.

Due to their inherent robustness and potential for disorder-resilient technologies, topological phases in synthetic periodic platforms are a very active area of study.
Notably, the initial demonstrations of higher-order topological phases \cite{Science357Benalcazar} were produced using synthetic materials \cite{Nature555Peterson, Nature555SerraGarcia, PRL124Qi, NatPhys14Imhof, NatPhoton13Mittal}.
A quadrupole topological insulator (QTI) is the first member of the multipole higher-order topological insulators, but is not straightforward to implement as it requires a $\pi$ flux of synthetic magnetic field threading each plaquette in the lattice~\cite{PhysRevB96Benalcazar, Science357Benalcazar}.
In previous experimental realizations of QTIs \cite{Nature555Peterson, Nature555SerraGarcia, PRL124Qi, NatPhys14Imhof, NatPhoton13Mittal}, the $\pi$ flux was achieved by tailored connectivities within the system \cite{Nature555Peterson,Nature555SerraGarcia,PRL124Qi, NatPhys14Imhof} or couplings with arbitrary phases by exploiting additional coupling links \cite{NatPhoton13Mittal}.
However, these approaches are not always practical, especially in nanoscale geometries, and a good solution is needed to enable synthetic fluxes in a broader range of experimental platforms.

\begin{figure}[htbp]
	\centering
	\includegraphics[width=12cm]{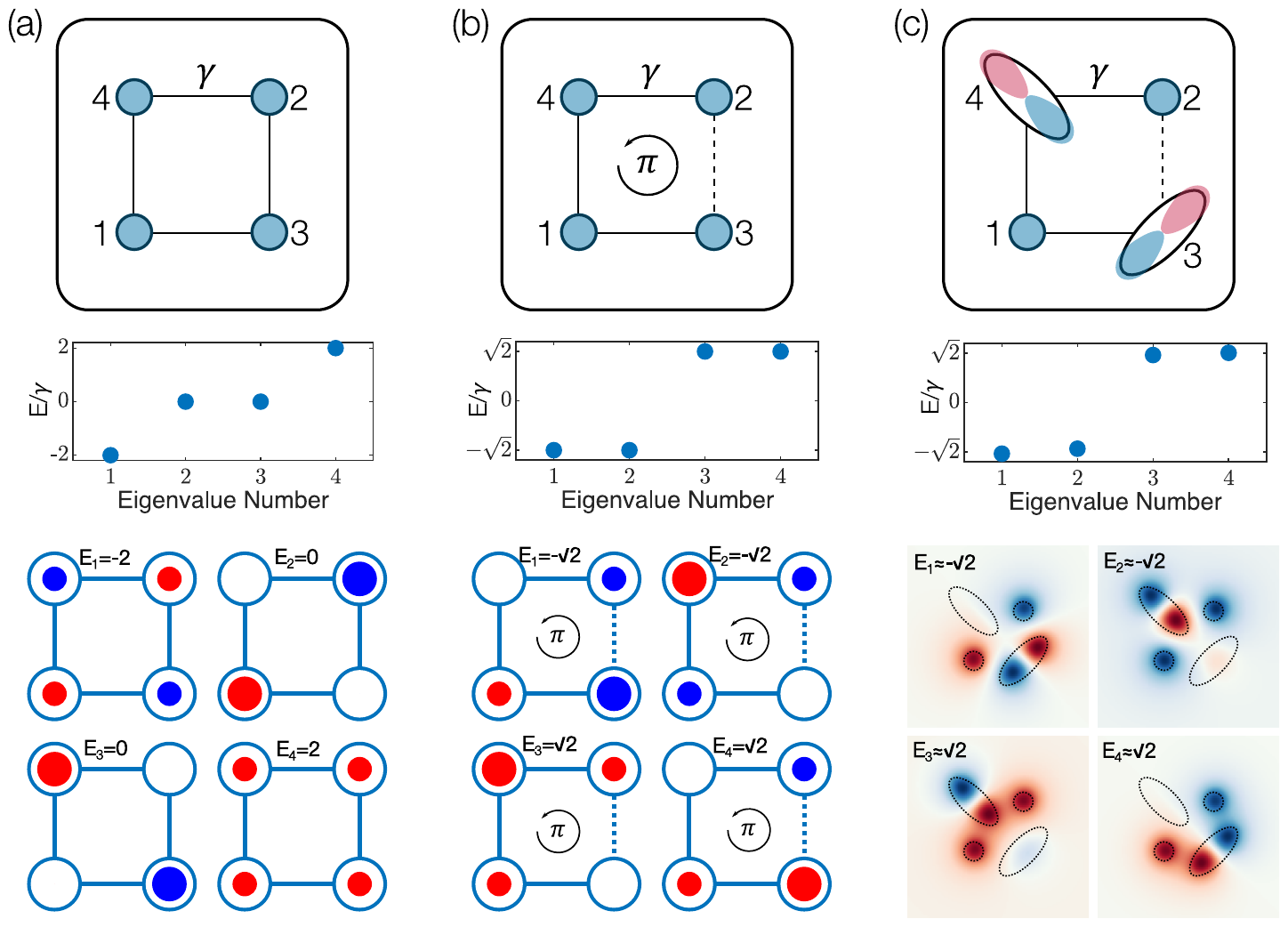}
	\caption{\textbf{Synthetic $\pi$ flux threading a unit cell plaquette induced by $s$ and $p$ orbitals.} (a)-(c) Schematic and corresponding eigenvalues and eigenmodes of a unit cell plaquette (a) composed of only $s$ orbitals without synthetic $\pi$ flux, (b) with synthetic $\pi$ flux,  and (c) composed of both $s$ and $p$ orbitals, respectively. The solid and dashed lines in the schematic indicate the positive and negative couplings, respectively. The areas and colors of the circles indicate amplitudes and phases of the corresponding eigenmodes, respectively. As can be seen in (a) and (b), the artificial introduction of a $\pi$ flux leads to a significant change in the eigenspectrum and eigenmodes. When a wavefunction in (c) crosses the $p$ orbital at site 3 a $\pi$-phase is accumulated, thereby inducing a flux in the plaquette. This effect is confirmed by the fact that the eigenvalues and eigenmodes from the full continuum calculations (c) line up with the states in (b).
	}
	\label{fig:UnitCellTB}
\end{figure}

Here, we present a QTI in a photonic system that uses the symmetry representations of on-site orbitals to generate the necessary synthetic $\pi$ fluxes.
We consider $s$ and $p$ orbitals, which have inherent even and odd parities, respectively.
We exploit the property that, as the wavefunctions of the collective system traverse a $p$ orbital, they accumulate a phase of $\pi$.
The combination and judicious control of the orbitals in a four-site unit cell [Fig.~\ref{fig:UnitCellTB}(c)], breaking the symmetries of mixed-orbital lattices, creates a synthetic $\pi$ flux that opens a gap at ``half-filling" which, along with the modulation in the hopping amplitudes, results in the gapped system having a QTI phase.
We experimentally demonstrate the photonic QTIs in a waveguide lattice fabricated using direct laser writing by showing the existence of mid-gap modes, which are localized at the corners of the lattice.
The fabrication of the waveguides by direct laser writing allows for unprecedented control over the waveguide parameters including both the cross-sections and the trajectories of the waveguides \cite{jorg_dynamic_2017,jorg_artificial_2020}, which was not straightforward in the conventional femtosecond direct laser writing technique \cite{JPhysB43Szameit}.

We start by presenting the implementation of our QTI by using $s$ and $p$ orbitals to induce $\pi$ flux threading a unit cell, taking advantage of the $\pi$-phase accumulated by the wavefunction as it crosses the $p$ orbital.
A square unit cell is composed of four sites with different orbitals: two $p$ orbitals, which have the major axes tilted from the $y$-axis by $\pm 45^{\circ}$, respectively, and two $s$ orbitals as shown in Fig.~\ref{fig:UnitCellTB}(c).
We first validate that the $\pi$ flux threads through this square unit cell by comparing three unit cell plaquettes as shown in Fig.~\ref{fig:UnitCellTB}: those composed of only $s$ orbitals without and with $\pi$ flux, and a unit cell plaquette composed of both $s$ and $p$ orbitals.

To compare the three unit cell plaquettes, we computed the eigenvalues and eigenmodes of the first two cases, having only $s$ orbitals, using the tight-binding Hamiltonians, and the other case, having both $s$ and $p$ orbitals, using the full-continuum wave equation.
For a unit cell plaquette without a $\pi$ flux [Fig.~\ref{fig:UnitCellTB}(a)], the eigenvalues are $-2\gamma,\,0,\,2\gamma$, where only the zero-energy states are two-fold degenerate and $\gamma$ is the coupling rate between the nearest-neighbor sites.
Whereas for a unit cell plaquette, in that a $\pi$ flux threading the plaquette is induced by introducing a negative coupling [Fig.~\ref{fig:UnitCellTB}(b)], the eigenvalues are $\pm\sqrt{2}\gamma$, each of which is two-fold degenerate.
On the other hand, the eigenvalues and eigenmodes of the unit cell plaquette with both $s$ and $p$ orbitals calculated using the full-continuum wave equation show a great resemblance with the unit cell with the synthetic $\pi$ flux [Fig.~\ref{fig:UnitCellTB}(c)], where the bulk gap is opened.
The similarity between these two unit cell plaquettes validates that the specific arrangement of having two $p$ orbitals in the same unit cell aligned at different angles induces the effective magnetic flux of $\pi$ per plaquette.

\begin{figure}[htbp]
  \centering
  \includegraphics[width=12cm]{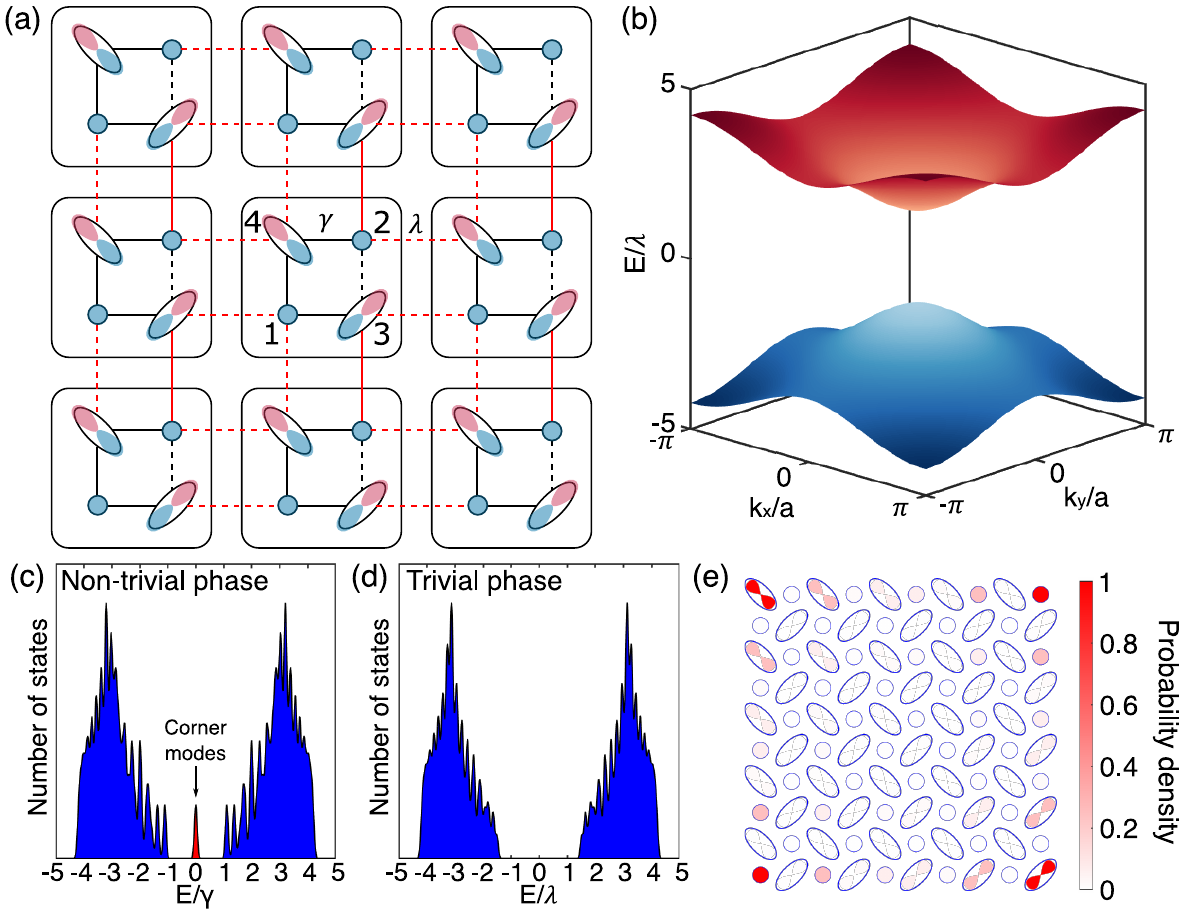}
  \caption{\textbf{Quadrupole topological insulator using orbital-induced synthetic flux.} (a) Schematic of the quadrupole topological insulator with orbital-induced synthetic flux. For the tight-binding model, $\gamma$ and $\lambda$ are the nearest-neighbor coupling terms within (black) and across (red) unit cells, respectively. Dashed lines represent coupling terms with negative signs due to the overlap with the ``negative part'' of the $p$ orbital. Numbers indicate the basis of the Hamiltonian. (b) A bulk band structure where $|\gamma/\lambda| = 1/2$. The band structure consists of two two-fold degenerate bands. (c) The numerically calculated density of states in the non-trivial phase ($|\gamma/\lambda|$=1/2) and (d) trivial phase ($|\gamma/\lambda|$=2), respectively, where the system has 10$\times$10 unit cells. (e) Combined eigenmode local density of states of the four topologically protected corner modes in the non-trivial phase ($|\gamma/\lambda|$=1/2). Here, the system has 5$\times$5 unit cells as in the experiment.}
  \label{fig:schematic}
\end{figure}

For the tight-binding description of the model, we choose a base where all $s$ orbitals have a phase of zero and one fixed site of the $p$ orbital has a phase of zero while the opposite site has a phase of $\pi$. 
In Fig.~\ref{fig:UnitCellTB}(c) and Fig.~\ref{fig:schematic}(a), zero and $\pi$ phases of the basis states are colored in blue and red, respectively.
As the hopping is determined by the overlap of the base states, the ``negative part'' of the $p$ orbital results in some negative hoppings.
Then, the tight-binding bulk Hamiltonian of this system becomes:

\begin{align}
	h^{oq}(\textbf{k},\delta) = &\left[\gamma-\lambda\cos(k_{x}a)\right]\Gamma_{4}+\lambda\sin(k_{x}a)\Gamma_{3}\nonumber\\
	&+\left[\gamma-\lambda\cos(k_{y}a)\right]\Gamma_{2}+\lambda\sin(k_{y}a)\Gamma_{1}+\delta \Gamma_{0},
	\label{eq:BlochHamiltonian}
\end{align}
where $a$ is the lattice constant, $\gamma$ and $\lambda$ are the nearest-neighbor coupling terms within and across unit cells, respectively, the $\Gamma$-matrices \cite{Science357Benalcazar} are $\Gamma_{j}=-\tau_{2}\bigotimes\sigma_{k}$ for $j\in\{1,2,3\}$, $\Gamma_{0}=\tau_{3}\bigotimes\sigma_{0}$, $\Gamma_{4}=\tau_{1}\bigotimes\sigma_{0}$ where $\tau$ and $\sigma$ are Pauli matrices for the degrees of freedom within a unit cell.
When the on-site energy of the $s$ and $p$ orbitals are the same, $\delta=0.$
This Hamiltonian closely resembles those Hamiltonians studied in the previous realizations of the quadrupole insulators \cite{Science357Benalcazar, Nature555Peterson, Nature555SerraGarcia}, where the only difference is the sign of $\lambda\cos(k_{x,y}a)$ in the first and third terms on the right-hand side.
The consequent difference is that at the phase transition point in the original model the bandgap closes at the $\textbf{M}$ point, while in our model it occurs at the $\boldsymbol{\Gamma}$ point.
In addition, this Hamiltonian has two mirror symmetries $M_{x}$ and $M_{y}$, which do not commute with each other, and also has $C_{4}$ symmetry.
Fig.~\ref{fig:schematic}(b) shows the 2D bulk band structure of our model in the topologically non-trivial phase, where the bandgap has opened at the $\boldsymbol{\Gamma}$ point due to the synthetic $\pi$ flux in the system.
For the system with open boundary conditions, as shown in Fig.~\ref{fig:schematic}(a), the non-trivial quadrupole phase leads to gapped edge states and in-gap corner states, as shown in Fig.~\ref{fig:schematic}(c-e) (see Supplementary Section~\ref{section:supp_quadrupole-phases} for a more detailed discussion of the symmetries and the topological quadrupole phases of the Hamiltonian).

We experimentally verify the quadrupole topology of the system by considering a two-dimensional lattice of evanescently-coupled waveguides.
A square unit cell is composed of four waveguides: two elliptical waveguides, which have the major axes tilted from the $y$-axis by $\pm\SI{45}{\degree}$, respectively, and two circular waveguides as shown in Fig.~\ref{fig:schematic}(a).
We control the radii of waveguides such that the lowest-energy mode (\smode) of the circular waveguides and the second-lowest-energy mode (\pmode) of the elliptical waveguides have the same propagation constant, enabling them to couple to one another.
The lowest-energy mode of the elliptical waveguides has a propagation constant far detuned from the other modes, such that they are well separated from our main system and can be neglected (see Supplementary Section~\ref{section:supp_full-continuum} for the Eigenmode calculations with the full-continuum Hamiltonian).

The radii of the major and minor axes of the elliptical waveguides in the experiment are \SI{0.6}{\micro\metre} and \SI{1.3}{\micro\metre}, and the radius of the circular waveguides is \SI{0.5}{\micro\metre}. 
The center to center distances between the waveguides determine the coupling strength, which in our structure are dimerized to be \SI{1.6}{\micro\metre} and \SI{2.1}{\micro\metre} for strong and weak couplings, respectively. 
The core of the waveguide is made out of the resin SU8-2 (Microchem) with a refractive index of $n_\text{core}=1.59$ and is surrounded by IP-Dip (Nanoscribe), which has a refractive index of $n_\text{clad}=1.54$.
The sample was fabricated using a Nanoscribe Photonic Professional GT \cite{jorg_dynamic_2017,fedorova_cherpakova_limits_2019,cohen_generalized_2020,APLPhoton6Schulz} (see Supplementary Section~\ref{section:supp_sample-fabrication} for Details about the Fabrication). For the measurements, light with a wavelength of \SI{760}{\nano\metre} from a white light laser (NKT photonics and a VARIA filter box) is injected with a 20$\times$ objective (NA=0.4) to a selected waveguide at the input facet of the waveguide array. 
After the propagation through the \SI{1}{\milli \metre} long structure, the diffracted light at the output facet is imaged with another 20$\times$ objective onto a CMOS camera (Thorlabs DDC1545M) [the measurement setup is sketched out in Fig.~\ref{fig:structure}(a)]. 
The equation governing the diffraction of light through the waveguide array is:
\begin{equation}
i\partial_{z}\psi(\textbf{r},z)=\hat{H}\psi(\textbf{r},z),
\label{eq:waveequation}
\end{equation}
where $\psi(\textbf{r},z)$ is the transverse electric field amplitudes at propagation distance $z$. $\hat{H}$ is the wavelength ($\lambda$) dependent continuum Hamiltonian for the wave propagation in the waveguide array.
Since we only consider a single bound mode for each waveguide and it evanescently couples to the neighboring waveguides, we can approximate the diffraction of light in our waveguide array using a tight-binding model, $i\partial_{z}\psi_{i}(z)=-\sum_{j}c_{ij}(\lambda)\psi_{j}(z)$, where $\psi_{i}$ is the amplitude in the $i$-th waveguide, and $c_{ij}(\lambda)$ is the coupling constant between waveguides $i$ and $j$ at wavelength $\lambda.$

\begin{figure}[htbp]
  \centering
  \includegraphics[width=12cm]{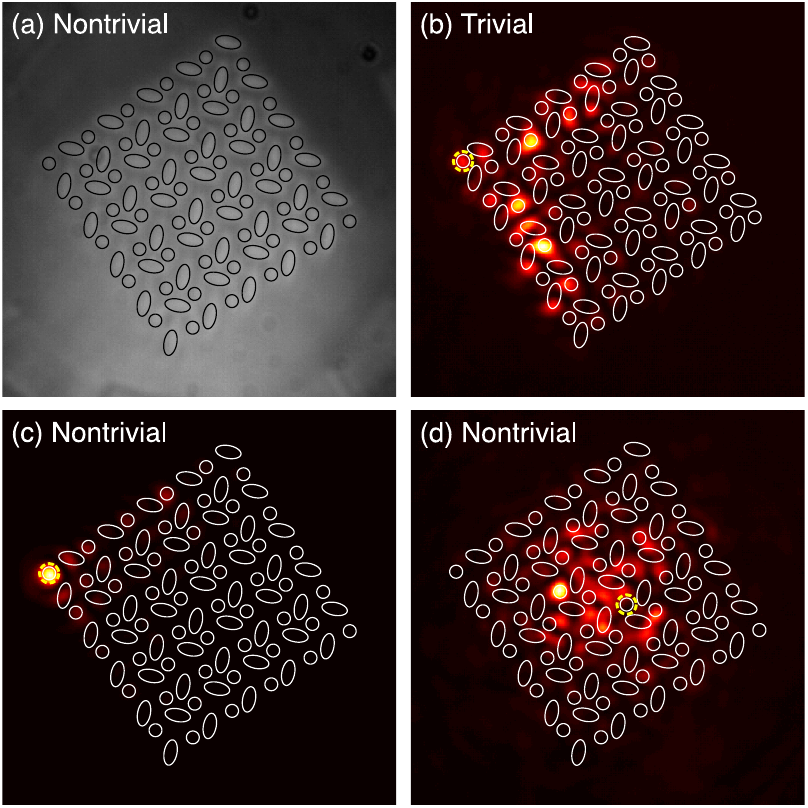}
  \caption{\textbf{Experimentally measured diffracted light at the output facet.} (a) Cross-sectional image of the output facet of the waveguide array in the non-trivial phase with a broad illumination of the input facet (see Supplementary Fig.~\ref{fig:structure}). (b-d) Measured intensity profiles at the output facet of the waveguide structures. Waveguides, where light is injected at the input facet, are indicated with yellow dashed circles. The intensity profiles are normalized to their respective maximum value to increase visibility. (b) Light is injected into the waveguide at the left corner of the waveguide array in the trivial phase and (c) non-trivial phase, respectively. (d) When light is injected into a waveguide in the center of the waveguide array in the non-trivial phase, it spreads into the bulk of the structure. In (a) and (b-d), black and white lines are overlapped to indicate the positions of the waveguides, respectively.}
  \label{fig:Exp_Direct}
\end{figure}

To experimentally observe the corner-localized topological modes, the light was injected into a waveguide at one of the corners of the waveguide array.
In Fig.~\ref{fig:Exp_Direct}(b-c), the diffracted light observed from the output facet for two different topological phases is shown.
When the waveguides located at the corners of the waveguide array are excited in a trivial phase, the injected light spreads significantly into the bulk [Fig.~\ref{fig:Exp_Direct}(b)], which indicates that there is no corner-localized eigenmode.
On the other hand, for the non-trivial phase, the light does not diffract into the bulk and is tightly confined close to the corner where the light was initially injected [Fig.~\ref{fig:Exp_Direct}(c)].
This confinement of the light at the injected corner is an indication of the presence of the corner modes and their localization is due to the non-trivial topology of the system.
To prove further that the spatial localization of the corner modes in the non-trivial phase is not due to the weak coupling between the waveguides, light is injected at a waveguide in the center of the waveguide array.
The injected light diffracts significantly into the bulk of the structure as shown in Fig.~\ref{fig:Exp_Direct}(d), which supports further that the corner-localized mode shown in Fig.~\ref{fig:Exp_Direct}(c) emerges due to the non-trivial topology of the model.

\begin{figure}[htbp]
  \centering
  \includegraphics[width=16cm]{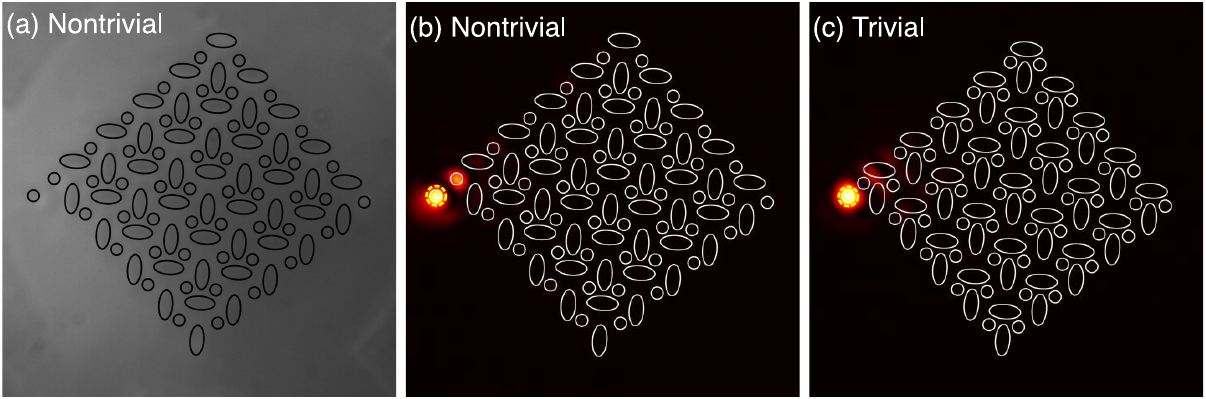}
  \caption{\textbf{Direct excitation of the corner mode using an auxiliary waveguide weakly coupled to the system.} (a) Cross-sectional image of the output facet of the waveguide array in the non-trivial phase with an auxiliary waveguide with a broad illumination of the input facet (see Supplementary Fig.~\ref{fig:structure}). (b) Diffracted light measured at the output facet when light is injected into the auxiliary waveguide directly at the left corner of the waveguide array in the non-trivial phase and (c) trivial phase, respectively. Waveguides, where light is injected at the input facet, are indicated with yellow dashed circles. In (a) and (b-c), black and white lines are overlapped to indicate the positions of the waveguides, respectively.}
  \label{fig:Exp_Aux}
\end{figure}

However, the localization of the corner modes alone does not prove the quadrupole properties, as this behavior was also observed in a system similar to ours, just without a $\pi$ flux.
In this system, however, the corner state lies not in a bandgap but is a bound state in the continuum \cite{PhysRevLett125Cerjan}.
To demonstrate that the corner localized modes in our system are indeed in a bandgap and topologically non-trivial modes due to the quadrupole topology, we introduce auxiliary waveguides as shown in Fig.~\ref{fig:Exp_Aux}(a).
The auxiliary waveguides are weakly coupled to the lattice such that they can be used as an external drive injecting light into the lattice at the energy of their bound modes without significantly altering the intrinsic modes of the lattice \cite{NatPhoton12Noh}.
In the experiment, the center to center distance from the auxiliary waveguide to the waveguide at the corner is \SI{2.1}{\micro\metre}.
The auxiliary waveguide is identical to the circular waveguides in the lattice, therefore the energy of the light injected from the auxiliary waveguide into the lattice is at zero energy.
In our system, the light initially injected at the auxiliary waveguide couples only to the corner state in the non-trivial phase [Fig.~\ref{fig:Exp_Aux}(b)] but does not couple into the system in the trivial phase [Fig.~\ref{fig:Exp_Aux}(c)]. 
This proves experimentally that in our system with $s$ and $p$ orbitals, the $\pi$ flux is induced in the unit cell, which opens the bandgap, and that the corner localized modes in the non-trivial phase are pinned at midgap due to the quadrupole topology of the system.

In this work, we have demonstrated synthetic crystalline structures composed of multiple orbitals.
We realized the quadrupole topological insulator in which synthetic $\pi$ flux threading each plaquette is induced due to the different symmetry representations of the orbitals.
To prove the non-trivial quadrupole topology of the system, we have experimentally verified that our realization of the quadrupole topological insulator has zero-dimensional corner-localized modes in the middle of the band gap.
Previously, the orbital degree of freedom has been nearly exclusively utilized in the ultracold atoms in optical lattices among diverse metamaterial systems.
Different experimental platforms have their unique features, such as abilities to more precisely or actively control the gain/loss, on-site energies, or coupling strengths, which can be associated with the orbital degree of freedom to open the possibility to study richer physics.
Furthermore, the realization of such photonic quadrupole topological insulators in the time-reversal symmetric system can provide a more straightforward route to utilize the quadrupole topology in practical applications, since the geometry we present is easier to implement than the one proposed previously.

\section*{Data availability}
The data that support the findings of this study are available from the corresponding author on reasonable request.

\section*{Acknowledgements}
We thank Christina J\"{o}rg for useful discussions. 
G.v.F. and J.S. acknowledge funding by the Deutsche Forschungsgemeinschaft through CRC/Transregio 185 OSCAR (project No.\ 277625399).
G.B. and J.N. acknowledge the support of the US Office of Naval Research (ONR) Multidisciplinary University Research Initiative (MURI) grant N00014-20-1-2325 on Robust Photonic Materials with High-Order Topological Protection. W.A.B. is thankful for the support of the Moore Postdoctoral Fellowship at Princeton University. 

\section*{Author contributions}
J.S., J.N. and W.A.B. conceived the project. J.S. and J.N. performed numerical simulations, designed the sample, and performed data analysis. J.S. fabricated and characterized the sample, and performed the optical probing experiment. G.B. and G.v.F. supervised all aspects of the project. All authors contributed in writing the manuscript.

\section*{Competing interests}
The authors declare no competing interests.

\section*{Additional Information}

\pagebreak
\widetext
\begin{center}
\textbf{\large Supplemental Materials: Photonic quadrupole topological insulator using orbital-induced synthetic flux}
\end{center}
\setcounter{equation}{0}
\setcounter{figure}{0}
\setcounter{table}{0}
\setcounter{page}{1}
\makeatletter
\renewcommand{\theequation}{S\arabic{equation}}
\renewcommand{\thefigure}{S\arabic{figure}}

\section{Quadrupole phases}
\label{section:supp_quadrupole-phases}

We discuss the bulk properties of the model by imposing the periodic boundary condition on a unit cell.
The symmetry group, that protects the quantization of both components of the polarization $p_x$ and $p_y$ and the quadrupole moment $q_{xy}$, includes two mirror symmetries $M_{x}=\tau_{3}\bigotimes\sigma_{1}$ and $M_{y}=\tau_{1}\bigotimes\sigma_{1}$ that do not commute with each other \cite{Science357Benalcazar}.
When $\delta=0$, the Hamiltonian (Eq.~\ref{eq:BlochHamiltonian}) has the required mirror symmetries,	$M_{x}h^{oq}(k_{x},k_{y})M_{x}^{\dagger} = h^{oq}(-k_{x},k_{y})$ and $M_{y}h^{oq}(k_{x},k_{y})M_{y}^{\dagger} = h^{oq}(k_{x},-k_{y})$.

These two mirror symmetries satisfy the condition for the quantization of the quadrupole moment such that they do not commute with each other and satisfy $\{M_{x}, M_{y}\}=0$.
In addition, the Hamiltonian (Eq.~\ref{eq:BlochHamiltonian}) has $C_{4}$ and chiral symmetries, $C_{4}h^{oq}(k_{x},k_{y})C_{4}^{\dagger} =h^{oq}(k_{y},-k_{x})$	and $\mathcal{C}h^{oq}(k_{x},k_{y})\mathcal{C}^{\dagger}=-h^{oq}(k_{x},k_{y})$, 
where $C_{4}=\left\{\left(\tau_{1}+i\tau_{2}\right)\bigotimes\sigma_{0}- \left(\tau_{1}-i\tau_{2}\right)\bigotimes i\sigma_{2}\right\}/2$ and $\mathcal{C}=\tau_{3}\bigotimes\sigma_{0}$.
The bulk Hamiltonian is gapped for $|\gamma/\lambda|\neq 1$ but closes at $\boldsymbol{\Gamma}$ point when $|\gamma/\lambda|=1$, where the topological transition occurs. Since the inversion symmetry $\mathcal{I}$ is related to the mirror symmetries as $\mathcal{I}=M_{y}M_{x}$, the Hamiltonian is also inversion symmetrical.

\begin{figure}[htbp]
  \centering
  \includegraphics[width=12cm]{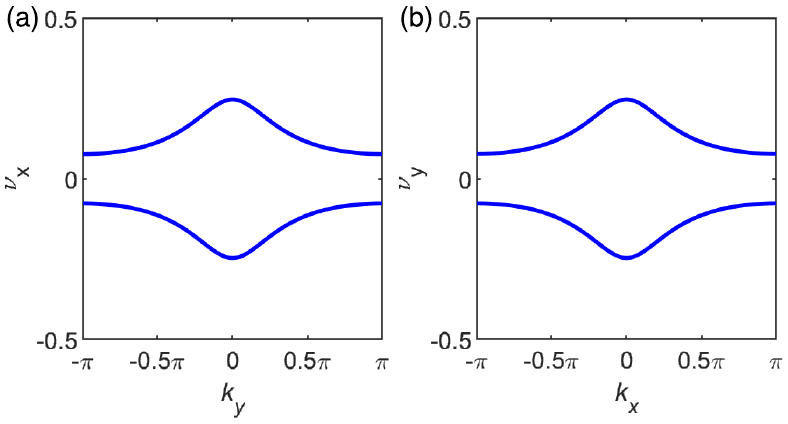}
  \caption{(a) Wannier bands $\nu_{x}(k_{y})$ and (b) $\nu_{y}(k_{x})$ computed for the bottom two bands of the bulk band structure where $\gamma/\lambda = 1/2$.}
  \label{fig:WannierCenter}
\end{figure}

We further study the quadrupole topology from the Bloch Hamiltonian by considering the Wannier bands and their polarization through the nested Wilson loop formulation \cite{Science357Benalcazar}.
The Wannier centers $\nu^{j}_{x}(k_{y})$ are proportional to the phases of the eigenvalues of the Wilson loop operator, $\mathcal{W}_{x}$ [Fig.~\ref{fig:WannierCenter}(a)], and the polarization $p_{x}$ can be obtained by taking the integral of the Wannier bands over the whole Brillouin zone in the $y$-direction.
The polarization $p_{y}$ can be determined similarly by considering $\mathcal{W}_{y}$ and the corresponding $\nu^{j}_{y}(k_{x})$ [Fig.~\ref{fig:WannierCenter}(b)].
Note that for the Bloch Hamiltonian in our model, the Wannier bands are gapped and $p_{x}=p_{y}=0$, which indicates that the bulk dipole moments are zero.
Here, we denote the upper and lower Wannier bands as $\nu_{x}^{\pm}$ ($\nu_{y}^{\pm}$), respectively.
Using the nested Wilson loop method, we can also compute the polarizations of each Wannier band.
The aforementioned existence of non-commuting mirror symmetries $M_{x,y}$ quantizes the Wannier-sector polarizations $p_{x}^{\nu_{\pm}}$ and $p_{y}^{\nu_{\pm}}$ to be either 0 or 1/2, which subsequently quantizes the bulk quadrupole $q_{xy}=2p_{x}^{\nu_{\pm}}p_{y}^{\nu_{\pm}}$ to be either 0 or 1/2.
From the calculation we find that $p_{x,y}=0$ for $|\gamma/\lambda|\neq 1$. On the other hand, we find that $q_{xy}=0$ for $|\gamma/\lambda|>1$ but $q_{xy}=1/2$ for $|\gamma/\lambda|<1$, which proves the quadrupole topology of the model.

\section{Eigenmode calculations using the full-continuum Hamiltonian}
\label{section:supp_full-continuum}

\begin{figure}[htbp]
	\centering
	\includegraphics[width=12cm]{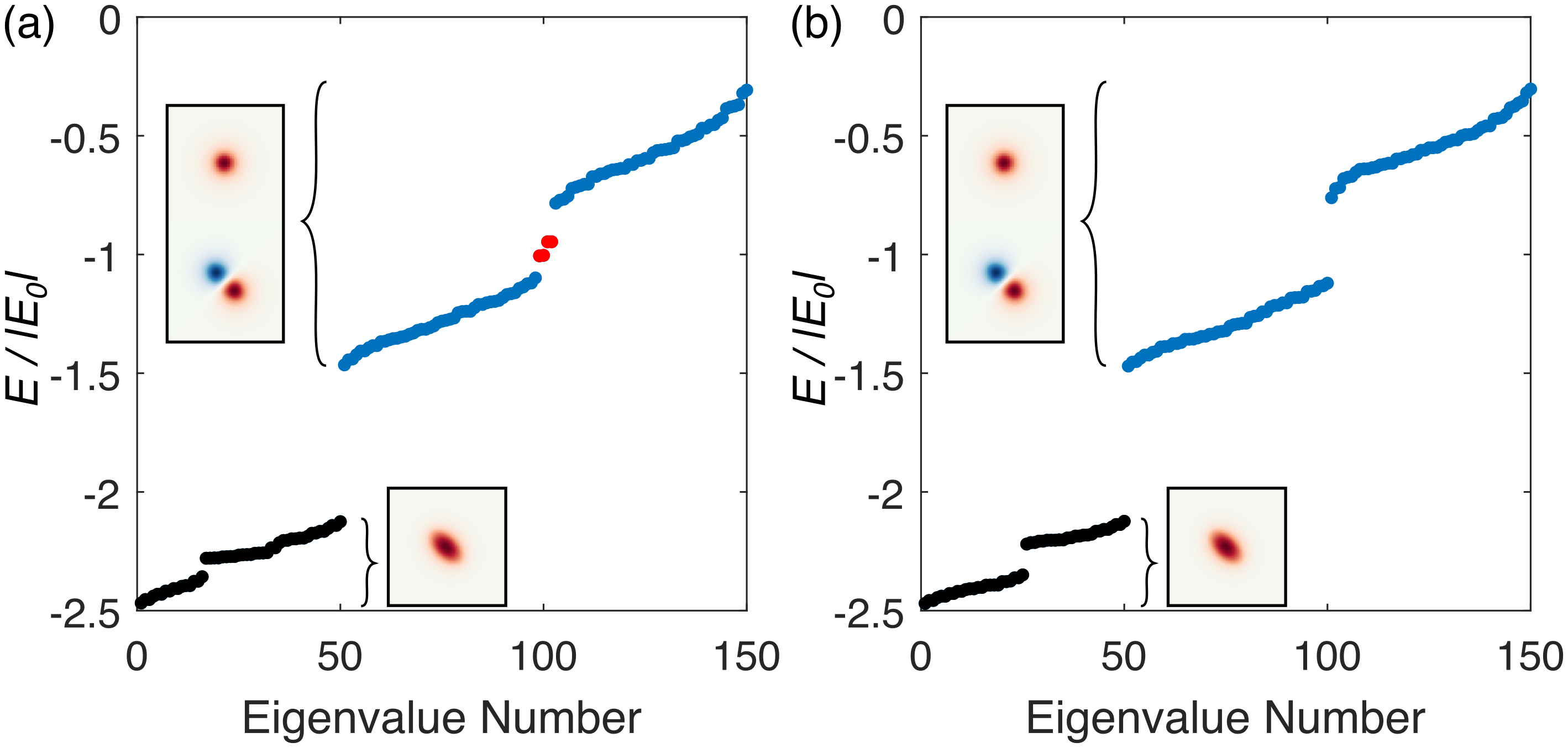}
	\caption{(a) Eigenenergies of the 5$\times$5 unit cell waveguide array in the non-trivial phase and (b) trivial phase, respectively. Eigenmodes are calculated using the full-continuum calculation by diagonalizing a continuum Hamiltonian for the propagation of the wave in the photonic lattice. Black dots indicate the bulk eigenmodes composed of the lowest-energy-mode of the elliptical waveguide. Blue(red) dots indicate the bulk(corner) eigenmodes composed of the lowest-energy mode of the circular waveguide (\smode) and the second-lowest-energy mode (\pmode) of the elliptical waveguides.}
	\label{fig:EigVal_All}
\end{figure}

While studying the evanescently-coupled waveguide system for realizing the photonic quadrupole topological insulators, we only considered the case where the lowest-energy mode (\smode) of the circular waveguides and the second-lowest-energy mode (\pmode) of the elliptical waveguides have the same energy.
The lowest-energy mode of the elliptical waveguide is ignored in the model since due to the judiciously controlled waveguide radii, the energy of this mode is well separated from the other modes of interest.
To confirm this, we calculated the eigenmodes using the full-continuum calculation by diagonalizing a continuum Hamiltonian for the propagation of the wave in the photonic lattice (Eq.~\ref{eq:waveequation}).
In Fig.~\ref{fig:EigVal_All}, we show the eigenmodes of the 5$\times$5 unit cell quadrupole topological insulator with orbital-induced synthetic flux, identical to the experiment.
As shown in Fig.~\ref{fig:EigVal_All}, the energy of the bulk band consisting of the lowest-energy-mode of the elliptical waveguide is spectrally well separated from the eigenmodes consisting of both \smode and \pmode.

\section{Details on the sample fabrication}
\label{section:supp_sample-fabrication}

\begin{figure}[htbp]
    \def\svgwidth{\columnwidth}
	\centering
	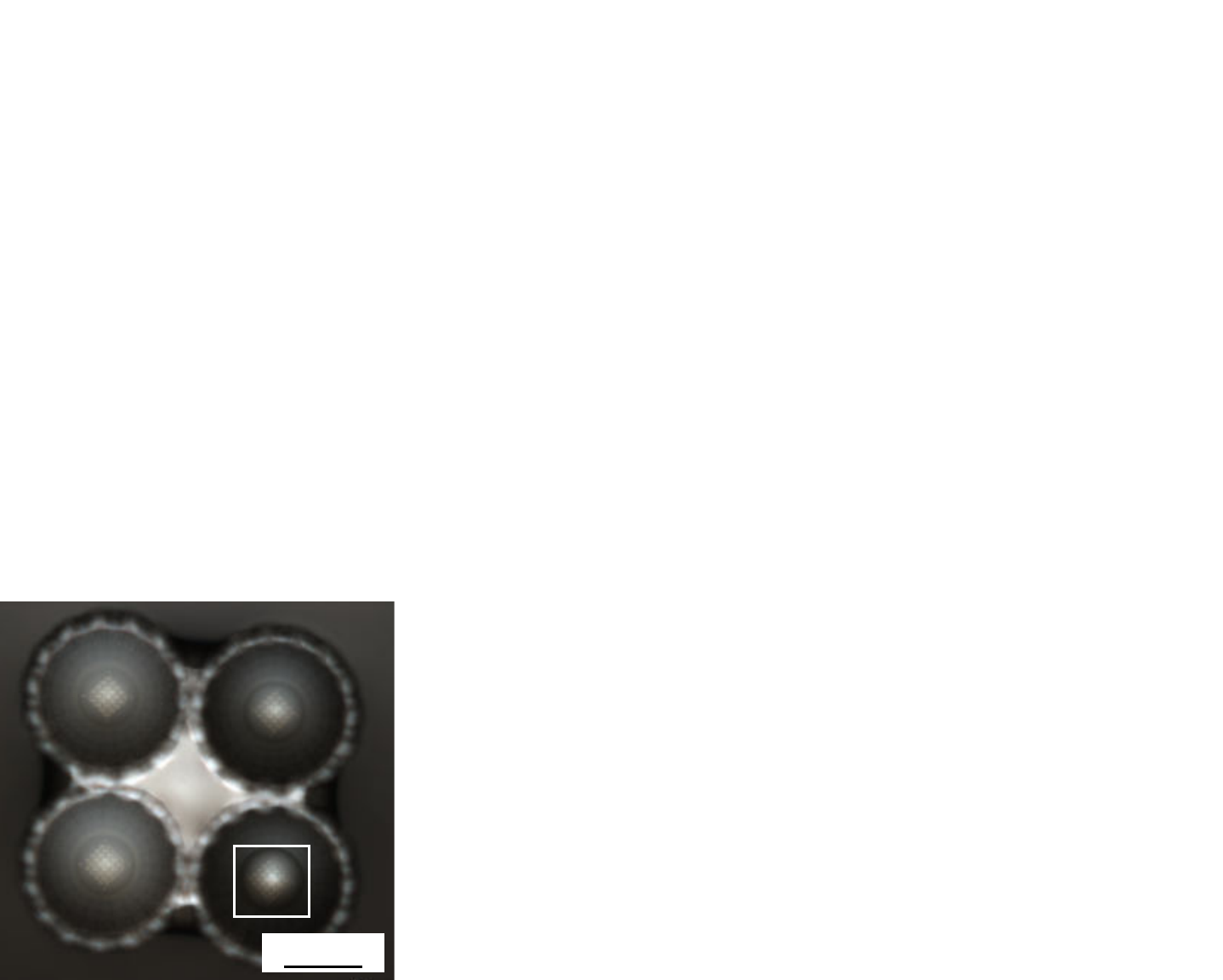
	\caption{(a) Schematic of the measurement setup. To image the input or output facet on the cameras a light-emitting diode is put into the beam path to serve as transmission illumination.
	(b) Microscope image of the input facet in reflection illumination of a final structure. Only after the baking process, the channels become waveguides and look brighter than the surrounding material.  
	(c) Microscope image of the input facet in transmission illumination of the waveguide array marked in (b) with a white box.
	(d) SEM Image of an inverse waveguide structure after development before the infiltration.
	}
	\label{fig:structure}
\end{figure}

The sample was fabricated using the Nanoscribe Photonic Professional GT similar to the way laid out in \cite{APLPhoton6Schulz}. 
First, the structure is 3D-printed by two-photon lithography in a negative-tone photoresist (IP-Dip, Nanoscribe). 
After the development (\SI{1}{\hour} in Propylene glycol methyl ether acetate and \SI{1}{\hour} in Isopropanol) one is left with the inverse waveguide structure like the one shown in Fig.~\ref{fig:structure}(d). 
A surrounding support grid is rotationally symmetrically designed and written with higher laser power to minimize distortion of the waveguide structure due to shrinkage of the photoresist during development. 
The next step is to dip the structure in \gbl. After one hour, where the structure is left to soak in the solvent, most of the \gbl is removed so that only the structure is wetted. 
A drop of SU8-2 (MicroChem) is then placed on the structure so that the SU8 can diffuse into the channels filled with \gbl. 
Finally, the sample is slowly heated up (\SI{10}{\kelvin\per\minute}) with a hotplate at \SI{150}{\degreeCelsius} for \SI{5}{\minute} to solidify the SU8.
The pre-infiltration with \gbl helps to increase the likelihood that the thin channels will be fully infiltrated, as the SU8 enters the channels by diffusion. 
Otherwise, the infiltration process would only rely on capillary forces, which can be hindered, e.g., by thin polymer threads. 
However, the SU8 should be diluted as little as possible with \gbl because it decreases in volume when baked out, allowing it to retract into the channels. 
After the baking process, the channels are filled with a material with a higher refractive index as IP-Dip and become waveguides and look brighter than the surrounding material [see Fig.~\ref{fig:structure}(b,c)].

\bibliography{reference}

\end{document}